\documentclass[conference]{IEEEtran}
\IEEEoverridecommandlockouts
\usepackage{cite}
\usepackage{amsmath,amssymb,amsfonts}
\usepackage{algorithmic}
\usepackage{graphicx}
\usepackage{textcomp}
\usepackage{xcolor}

\usepackage{bm}
\usepackage{algorithm}
\usepackage{multirow}
\usepackage{subfigure}
\usepackage{multirow}
\usepackage{romannum}
\usepackage{booktabs}
\usepackage[bookmarks=false]{hyperref}

\def\BibTeX{{\rm B\kern-.05em{\sc i\kern-.025em b}\kern-.08em
    T\kern-.1667em\lower.7ex\hbox{E}\kern-.125emX}}

\begin{document}

\title{Segmented Pairwise Distance for Time Series with Large Discontinuities}

\author{Jiabo He\textsuperscript{1}, Sarah Erfani\textsuperscript{1}, Sudanthi Wijewickrema\textsuperscript{2}, Stephen O’Leary\textsuperscript{2}, Kotagiri Ramamohanarao\textsuperscript{1}\\
\textsuperscript{1}School of Computing and Information Systems, The University of Melbourne, Australia\\
\textsuperscript{2}Department of Otolaryngology, The University of Melbourne, Australia\\
Email: \{jiaboh@student., sarah.erfani@, sudanthi.wijewickrema@, sjoleary@, kotagiri@\}unimelb.edu.au
}

\maketitle

\begin{abstract}
Time series with large discontinuities are common in many scenarios. However, existing distance-based algorithms (e.g., DTW and its derivative algorithms) may perform poorly in measuring distances between these time series pairs. In this paper, we propose the segmented pairwise distance (SPD) algorithm to measure distances between time series with large discontinuities. SPD is orthogonal to distance-based algorithms and can be embedded in them. We validate advantages of SPD-embedded algorithms over corresponding distance-based ones on both open datasets and a proprietary dataset of surgical time series (of surgeons performing a temporal bone surgery in a virtual reality surgery simulator). Experimental results demonstrate that SPD-embedded algorithms outperform corresponding distance-based ones in distance measurement between time series with large discontinuities, measured by the Silhouette index (SI).
\end{abstract}


\section{Introduction}
Time series are a ubiquitous form of data in scientific disciplines. There may be value gaps in time series, which are jumps of value orthogonal to the time axis. Time series with large discontinuities (value gaps) are common in a variety of scenarios, such as surgical procedures, human activity, etc. It is quite challenging to measure distances between these time series with the existence of such large discontinuities. Large discontinuities can impede putting local characteristics into the spotlight. Since distance measurement for time series is the core of similarity analysis, classification and clustering, we should address this issue of measuring distances between time series with large discontinuities.

There are a large quantity of algorithms measuring distances between time series, among which Euclidean distance and dynamic time warping (DTW) along with their derivative algorithms are the most widely used. Many classification and clustering algorithms are based on Euclidean distance when all elements have the same dimension or length \cite{cunningham2007k, ester1996density, arthur2007k, wang2019dbsvec}. However, it can perform poorly when there is distortion in time series along the time axis \cite{mueen2016extracting}. DTW is also widely used for global distance measurement of time series, applied in a diverse range of domains including gesture recognition \cite{barczewska2013comparison, plouffe2016static}, time series classification \cite{wegner2019dtwsat}, trajectory clustering \cite{atev2010clustering}, disease detection \cite{varatharajan2017wearable}, etc. DTW can address distortion in time series to certain extent by aligning two time series with indices in monotonically increasing order. DTW is a global distance-based algorithm which cannot fully extract local characteristics of time series. Due to this, DTW is unsuitable for certain kinds of data where local similarity is more significant than global similarity.

\begin{figure}[pt]
    \centering
    \subfigure[DTW]
    {
    \includegraphics[width=0.22\textwidth]
    {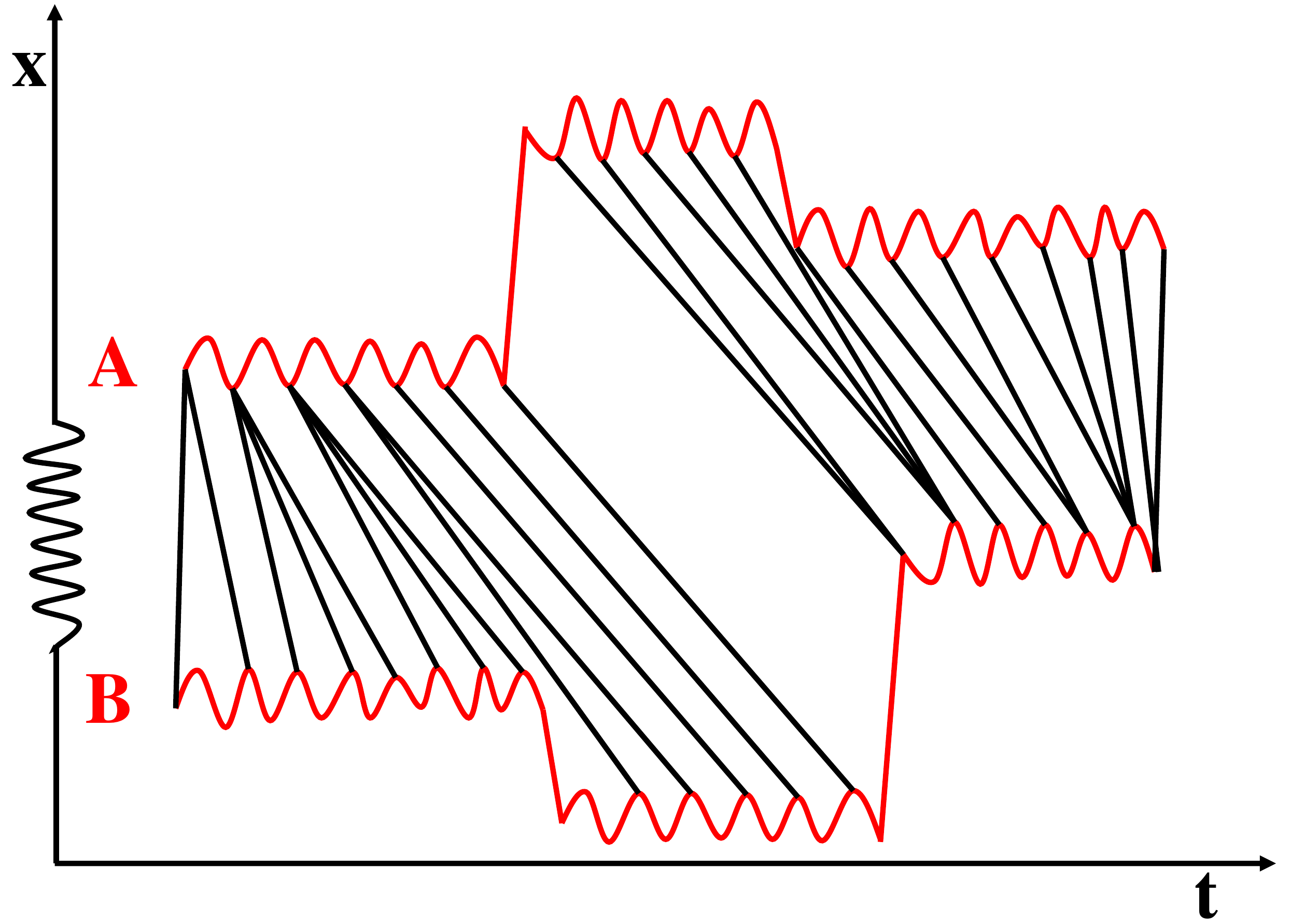}
    \label{image:example1}
    }
    \subfigure[SDTW]
    {
    \includegraphics[width=0.22\textwidth]{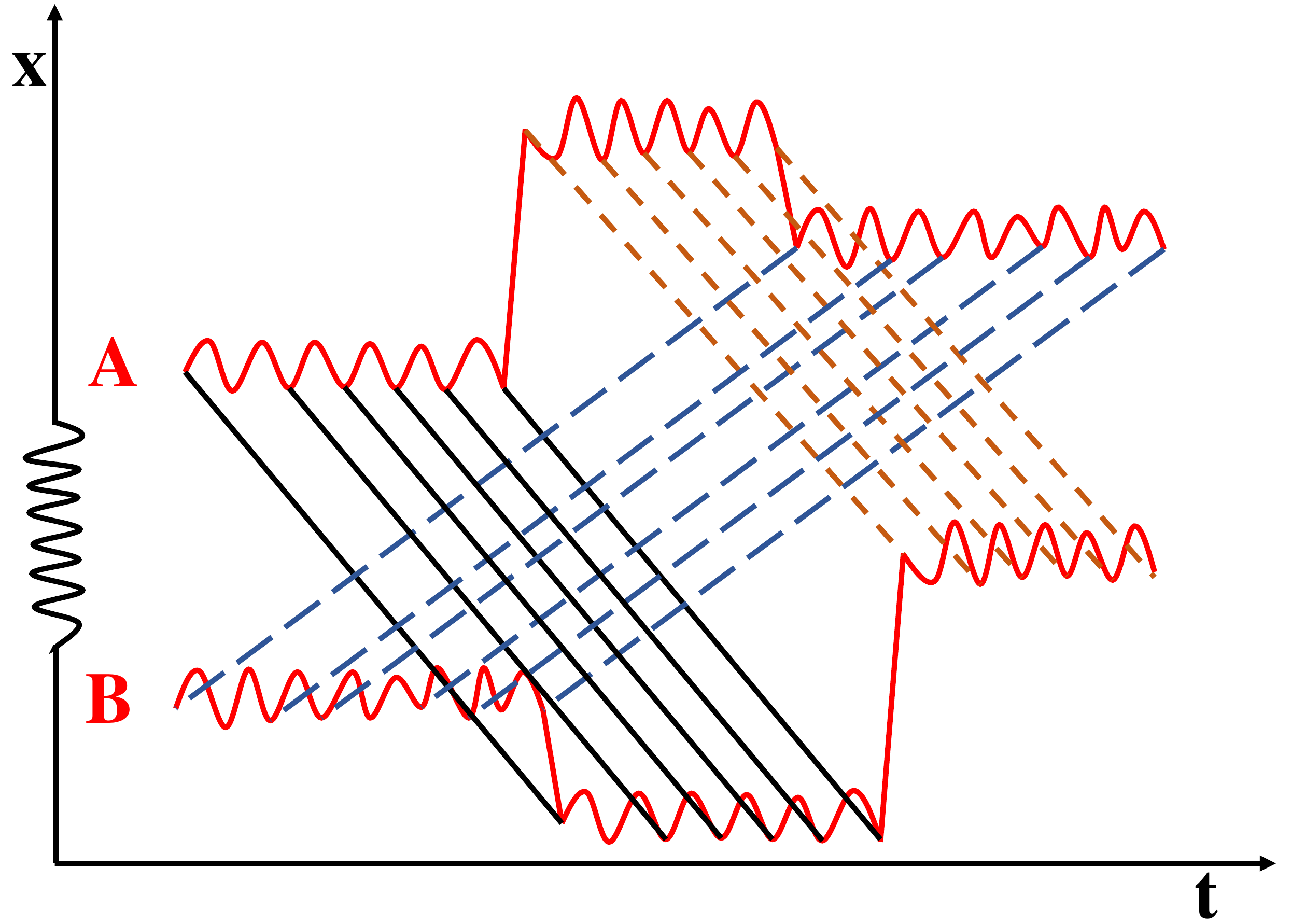}
    \label{image:example2}
    }
    \caption{Example of segmentation and alignment for time series with large discontinuities using DTW and SPD-embedded DTW (SDTW).}
    \label{image:example}
\end{figure}

In order to address the above issue, we propose the segmented pairwise distance (SPD) algorithm, which can be embedded in distance-based algorithms. Although both Euclidean distance and DTW have difficulty in measuring distances between time series with large discontinuities, we use DTW and its derivative algorithms as our baselines,  since DTW performs comparatively better than Euclidean distance. Fig. \ref{image:example} is an example to exhibit the difference between DTW and SPD-embedded DTW (SDTW) when dealing with time series with large discontinuities. DTW aligns two time series with indices monotonically increasing, while SDTW can segment time series based on large discontinuities and sum all distances between the most similar segment pairs. As a result, SDTW is able to detect similar time series sharing similar segment pairs by obtaining small overall distances between them.

This paper mainly contributes to proposing a new algorithm, SPD, to measure distances between time series with large discontinuities. Moreover, there are two technical merits in SPD: \textbf{(1)} SPD is orthogonal to distance measurement and can be embedded in all distance-based algorithms; \textbf{(2)} SPD can decide a unique segmentation threshold for every time series in different datasets so that it can be applied to a variety of datasets. We validate advantages of SPD-embedded algorithms over corresponding distance-based ones on both open datasets and the surgical dataset that we collect, where surgeries are performed by expert surgeons on the same patient's temporal bone in the virtual reality surgery simulator. It will provide a new challenging benchmark dataset for distance measurement of time series with large discontinuities\footnote{The code and dataset is available at \url{https://github.com/Jacobi93/Segmented-Pairwise-Distance}.}. In addition, all techniques for speeding up distance-based algorithms can also be applied to SPD-embedded algorithms, which is beyond the scope of this paper.

\section{Related Work}
\label{section2}

\subsection{Segmentation algorithms for time series}
Many segmentation algorithms for time series are based on regression subroutines \cite{levy2008catching, lovric2014algoritmic, wu2018finding}. There are three main categories of regression-based segmentation algorithms, including Sliding-Windows, Top-Down, and Bottom-Up algorithms. They are utilized to address multiple segmentation problems: generate piecewise representations (1) using only K segments, (2) minimizing the piecewise error, (3) and minimizing the total error in various scenarios \cite{keogh2004segmenting}. \cite{levy2008catching} caught segmentation points by combining the Lasso penalty with dynamic programming. \cite{han2015time} first learned a sequence of local relationship models that could best fit time series data, and then combined changes of local relationships to identify the operational behavior switching in the system level. Regression-based segmentation algorithms have two main limitations: (1) regression subroutines are not efficient if one just needs segments of time series without regression; (2) linear representations for univariate segments are not applicable in high-dimensional multivariate time series.

There are also some segmentation algorithms for time series without employing regression. \cite{graves2009multivariate} successfully segmented multivariate time series with differential evolution. Later, the fast low-cost online semantic segmentation (FLOSS) algorithm segmented time series at a high level by detecting the regime change \cite{gharghabi2017matrix}. In addition, Matrix Profile distance (MPdist) was proposed to detect the similarity of two time series when they share multiple similar subsequences based on Euclidean distance \cite{gharghabi2018ultra}. However, these algorithms still have some limitations: (1) they are not computationally efficient because they traverse all possible subsequences; (2) there are some hyperparameters (e.g., the length of subsequences in time series and the quantile threshold in MPdist) to be decided before measurement, which is domain dependent.

\subsection{Euclidean distance, DTW and their derivatives}
Both Euclidean distance and DTW have difficulty in measuring distances between time series with large discontinuities. Euclidean distance regards each time series as a high-dimensional point, which is extensively applied as the distance function in time series classification \cite{cunningham2007k}, clustering \cite{ester1996density, arthur2007k}, and other scenarios. In addition, DTW and its derivative algorithms can also measure distances between time series. Complexity invariance was proposed to measure distances between time series with varying
complexities, which could be embedded in DTW as the complexity-invariant DTW (CIDTW) algorithm \cite{batista2011complexity}. The shape of time series is also a significant feature. \cite{keogh2001derivative} proposed the derivative DTW (DDTW) algorithm to align time series using high level features of shape. Moreover, the phase difference is also a potential problem because DTW provides non-linear alignments. \cite{jeong2011weighted} proposed the weighted DTW (WDTW) algorithm to penalize points with higher phase difference, in order to achieve minimum distance distortion caused by outliers. The weighted version of DDTW (WDDTW) was then proposed in \cite{jeong2011weighted}. These derivative algorithms of DTW all perform competitively well in their specific scenarios and exhibit their limitations in others. They are all whole time series distance-based algorithms, which measures distances between time series using all elements in them. We do not consider those shapelet-based, interval-based, or dictionary-based algorithms. They measure distances between subsequences from whole time series with different feature selection methods, which were compared and analyzed comprehensively in \cite{bagnall2017great}.

Some research embedded segmentation techniques in DTW for distance measurement as well. \cite{ma2018segmentation} implemented peak identification and pairing for time series before DTW analysis. The limitation is that the number of segments between two time series must be the same, which is not ubiquitous in many scenarios. In contrast, our proposed segmented pairwise distance (SPD) algorithm can be embedded in any distance-based algorithm and the numbers of segments of two time series are not restricted to be the same. Another segmented-based DTW (SBDTW) algorithm was proposed for similarity measurement in urban transportation systems \cite{mao2017segment}. Point-segment, prediction and segment-segment distances were defined in SBDTW. The minimal distance of time series pairs was computed by accumulating the minimum of three distances. Instead, SPD only segments time series based on their large discontinuities. Following that, original distance-based algorithms are employed on every segment pair from different time series to obtain pairwise distances, which is then used for calculating the overall distance between time series.

\subsection{Local similarity}
Some researchers also noticed the significance of local similarity for time series. Local descriptors for recognizing motion patterns in videos were presented to classify human actions \cite{laptev2004local}. Internal self-similarities were captured by a local self-similarity descriptor \cite{shechtman2007matching}, which provided matching capabilities of complex visual data. Besides, all-pairs-similarity-search algorithms were also proposed to evaluate similarity joins for time series subsequences \cite{yeh2016matrix}. To extend this idea, our work measures local similarity between time series with large discontinuities using SPD-embedded algorithms.

\section{Segmented Pairwise Distance Algorithm}
\label{section3}

This paper proposes the segmented pairwise distance (SPD) algorithm to measure distances between time series with large discontinuities. The SPD algorithm can be embedded in all distance-based ones. Since DTW performs comparatively better than Euclidean distance, we use DTW and its derivative algorithms in our experiments. We embed SPD in DTW to build the SDTW algorithm as an example, with details in Algorithm \ref{algo:SDTW}.

\subsection{Dynamic Time Warping}
Before we go deep into our proposed SPD, it is necessary to introduce DTW first. We select DTW and its derivative algorithms as distance measurement baselines because DTW performs comparatively better than Euclidean distance in measuring distances between time series with large discontinuities. DTW is flexible to align time series with variant lengths. Equations \eqref{equ:1} and \eqref{equ:2} are recursive functions of DTW \cite{mueen2016extracting}. $\bm{A}$ and $\bm{B}$ are two time series in sequence $(a_1, a_2, \dotsc, a_{n_1})$ and $(b_1, b_2, \dotsc, b_{n_2})$. $n_1$ and $n_2$ are the number of elements in $\bm{A}$ and $\bm{B}$. $d(a_i,b_j)$ is the distance (usually it is Euclidean distance) defined between the $i^{th}$ element in $\bm{A}$ and the $j^{th}$ element in $\bm{B}$. 
\begin{equation}
\begin{split}
  D(1{:}n_1,1{:}n_2) = d(a_{n_1},b_{n_2}) + min[D(1{:}(n_1-1), \\ 1{:}(n_2-1)), D(1{:}(n_1-1),1{:}n_2),D(1{:}n_1,1{:}(n_2-1))]
\end{split}
\label{equ:1}
\end{equation}
\begin{equation}
  D(1{:}1,1{:}1) = d(a_1,b_1)
  \label{equ:2}
\end{equation}

\subsection{SPD-embedded Dynamic Time Warping}

\begin{algorithm}[tbp]
    \begin{flushleft}
            \textbf{Input:} Time series $\bm{A}$ and $\bm{B}$ in length $n_1$ and $n_2$, $q$\\
            \textbf{Output:} $SDTW(\bm{A}, \bm{B})$
    \end{flushleft}
\begin{algorithmic}[1]
\caption{SDTW}
\label{algo:SDTW}
    \STATE Calculate consecutive distances for $\bm{A}$ and $\bm{B}$, obtain segmentation thresholds based on $q$ and sorted distance distributions of $\bm{A}$ and $\bm{B}$, respectively
    \STATE Segment $\bm{A}$ and  $\bm{B}$ into $s_1$ segments ($\bm{a_1}, \bm{a_2}, \dotsc, \bm{a_{s_1}}$), and $s_2$ segments ($\bm{b_1}, \bm{b_2}, \dotsc, \bm{b_{s_2}}$) based on thresholds
    \STATE Calculate $DTW(\bm{a_i}, \bm{b_j})$ for all $i$ in $[1, s_1]$ and $j$ in $[1, s_2]$ to obtain the DTW matrix \textbf{D} in size of $(s_1, s_2)$
    \STATE $Dis_1=\sum_{i=1}^{s_1}{min(row(i))}$
    \STATE Record the column numbers of $min(row(i))$ in step $4$ and delete recorded columns with $s'_2$ columns remained in \textbf{D'}
    \STATE $Dis_1=Dis_1 + \sum_{j=1}^{s'_2}{min(col(j))}$
    \STATE $\textbf{D}=\textbf{D}^T$, repeat steps 4-6 to obtain $Dis_2$
    \STATE $SDTW(\bm{A}, \bm{B}) = \frac{min(Dis_1, Dis_2)}{n_1 + n_2}$\\
\end{algorithmic}
\end{algorithm}

We are finally in the position to introduce the core contribution of our work. This paper proposes SPD to help distance-based algorithms measure distances between time series with large discontinuities. We embed SPD in DTW to build SDTW as an example (Algorithm \ref{algo:SDTW}). In order to measure the SDTW distance between two time series, we first calculate distances of consecutive elements for each one of them, respectively. The quantile $q$ of sorted distance distribution decides the segmentation threshold for each time series (step 1). The quantile $q$ is not completely domain agnostic. The knowledge of the dataset can help set a reasonable $q$ for distance measurement between time series, although we find that it is insensitive in range of $[0.9, 0.99]$ in most scenarios by experiments. $q=0.99$ represents that the time series will be segmented where the distance between two consecutive elements is larger than $99\%$ of all in the time series. For example, if there are about one thousand elements in the time series, it will be segmented into $10$ subsequences approximately. There is a unique segmentation threshold for every time series when $q$ is determined.

After segmentation of two time series based on their thresholds we obtain (step 2), we can calculate the DTW matrix for every segment pair from two different time series based on the DTW algorithm (step 3). Afterwards, all minimal distances from every segment in $\bm{A}$ to any segment in $\bm{B}$ are found and accumulated (step 4). There are probably some segments in $\bm{B}$, which are never paired by $\bm{A}$ in step 4. Then minimal distances from those remaining segments in $\bm{B}$ to any segment in $\bm{A}$ are then found and accumulated to obtain $Dis_1$ (steps 5-6). $Dis_2$ can be measured after transposing \textbf{D} and repeating steps 4-6 (step 7). Finally, we can obtain the SDTW distance between two time series $\bm{A}$ and $\bm{B}$ in step 8.

\begin{figure}[ptb]
    \centering
    \subfigure[DTW=22]
    {
    \includegraphics[width=0.22\textwidth]
    {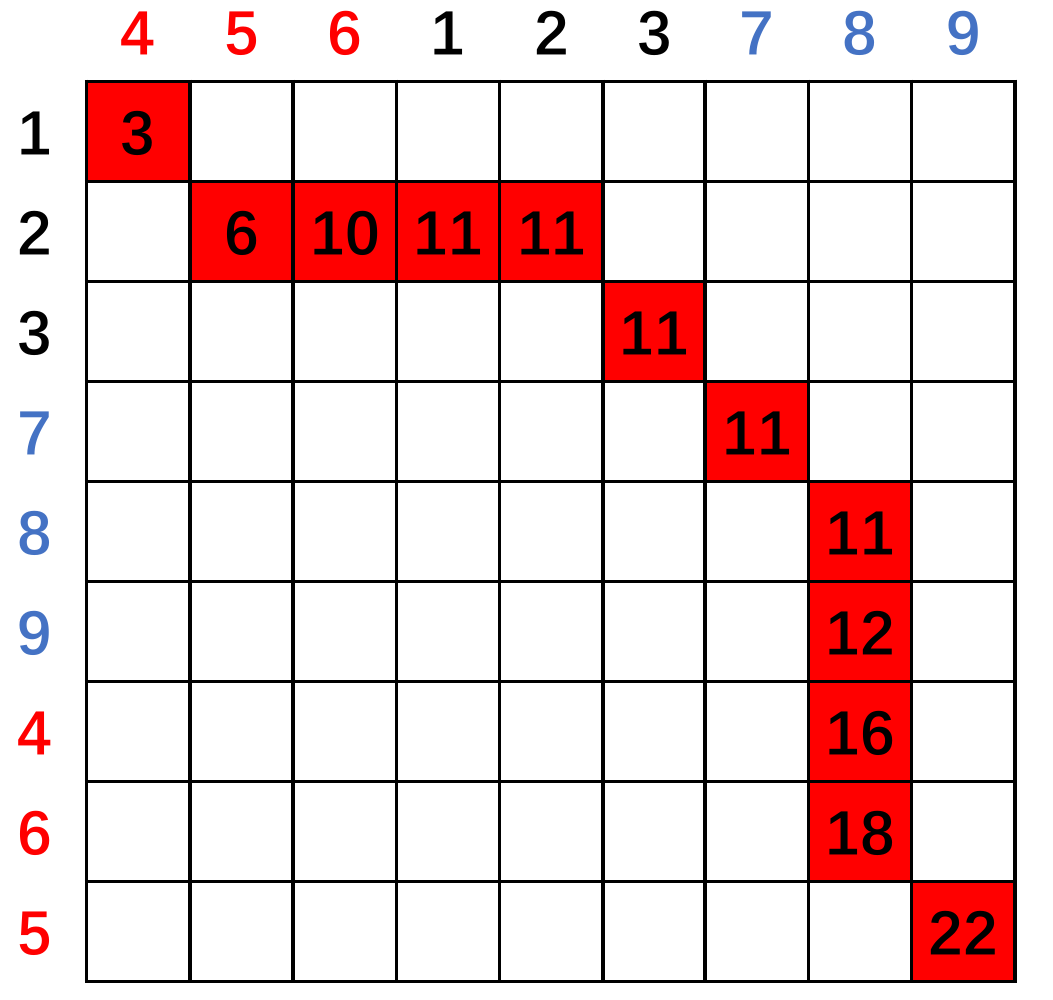}
    }
    \subfigure[SDTW=2]
    {
    \includegraphics[width=0.22\textwidth]{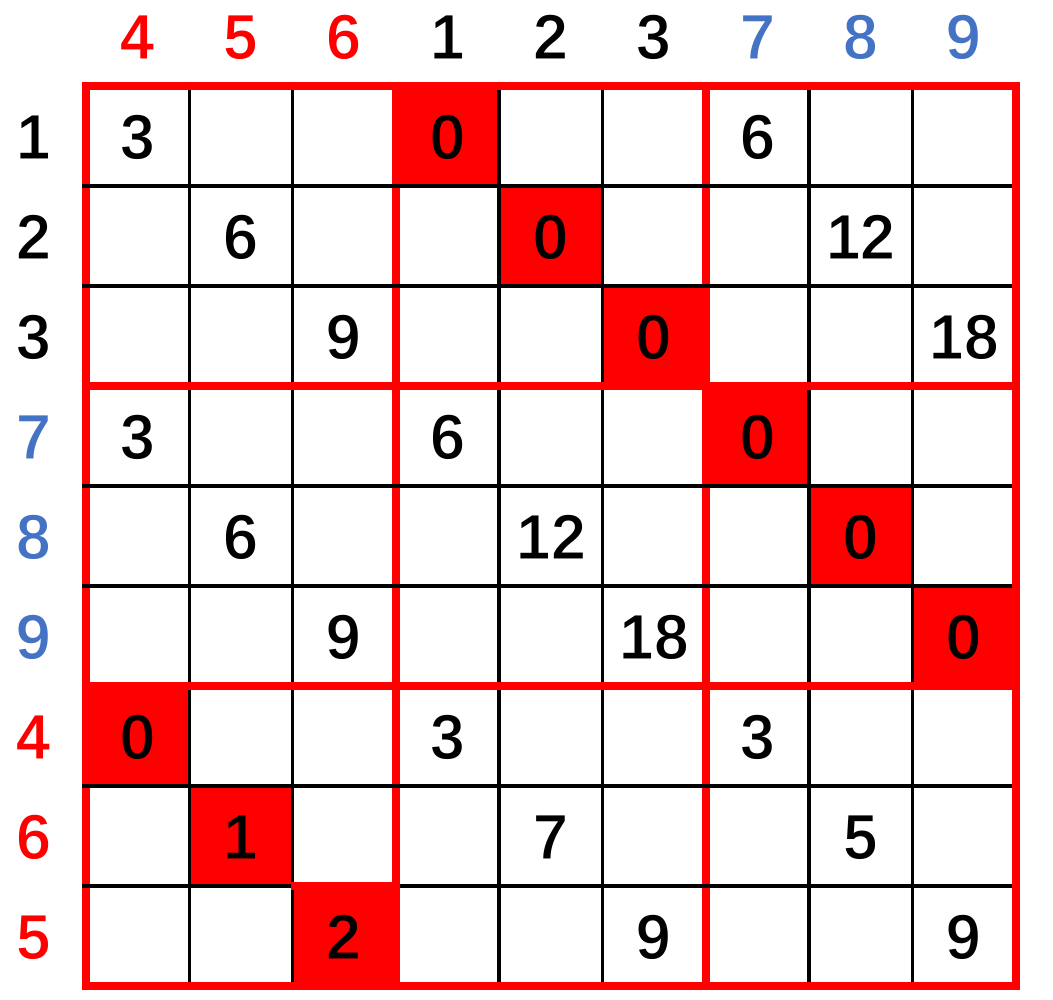}
    }
    \caption{Example of distance measurement between time series with large discontinuities using DTW and SPD-embedded DTW (SDTW).}
    \label{image:two}
\end{figure}

Here is an example of the comparison between DTW and SDTW when calculating the distance between two time series with large discontinuities (Fig. \ref{image:two}). Sequence $\bm{A}$ is (4, 5, 6, 1, 2, 3, 7, 8, 9). Sequence $\bm{B}$ is (1, 2, 3, 7, 8, 9, 4, 6, 5). We set the segmentation threshold to be 2 for each one of them so that each time series can be segmented into 3 subsequences. From the local point of view, two time series are very similar to each other but not exactly the same. Both $\bm{A}$ and $\bm{B}$ have subsequences (1, 2, 3) and (7, 8, 9) while there is a unique subsequence (4, 5, 6) in $\bm{A}$ and (4, 6, 5) in $\bm{B}$. SDTW outperforms DTW when analyzing local similarity between two time series, which successfully detects similar subsequence pairs from different time series.

The time complexity of DTW is $O(n_1n_2)$. The time complexity of SDTW is $O(n_1)+O(n_2)+O(\sum_{i=1}^{s_1}\sum_{j=1}^{s_2}l(\bm{a_i})l(\bm{b_j}))+O(s_1s_2)$, where $O(n_1)+O(n_2)$ is the time complexity for segmenting $\bm{A}$ and $\bm{B}$ (steps 1-2), $O(\sum_{i=1}^{s_1}\sum_{j=1}^{s_2}l(\bm{a_i})l(\bm{b_j}))$ is that for calculating the DTW matrix (step 3) and $O(s_1s_2)$ is that for calculating the SDTW distance based on the DTW matrix (steps 4-8). $l(\bm{a_i})$ is the length of the $i^{th}$ subsequence in $\bm{A}$ and $l(\bm{b_j})$ is the length of the $j^{th}$ subsequence in $\bm{B}$. Since $\sum_{i=1}^{s_1}l(\bm{a_i})=n_1$ and $\sum_{j=1}^{s_2}l(\bm{b_j})=n_2$, $O(\sum_{i=1}^{s_1}\sum_{j=1}^{s_2}l(\bm{a_i})l(\bm{b_j}))$ is equal to $O(n_1n_2)$. Because $O(s_1s_2)<O(n_1n_2)$, the time complexity of SDTW is $O(n_1n_2)$ in the end, which is the same as that of DTW although it takes a little longer to implement SDTW than DTW in experiments. In addition, all techniques for speeding up DTW can also be applied to SDTW, and SDTW has the advantage of less space complexity over DTW. Both of them are beyond the scope of this paper.

\section{Datasets}
\label{section4}


\subsection{The Cortical Mastoidectomy dataset}
\begin{figure}[tbp]
  \centering
  \includegraphics[width=5cm]{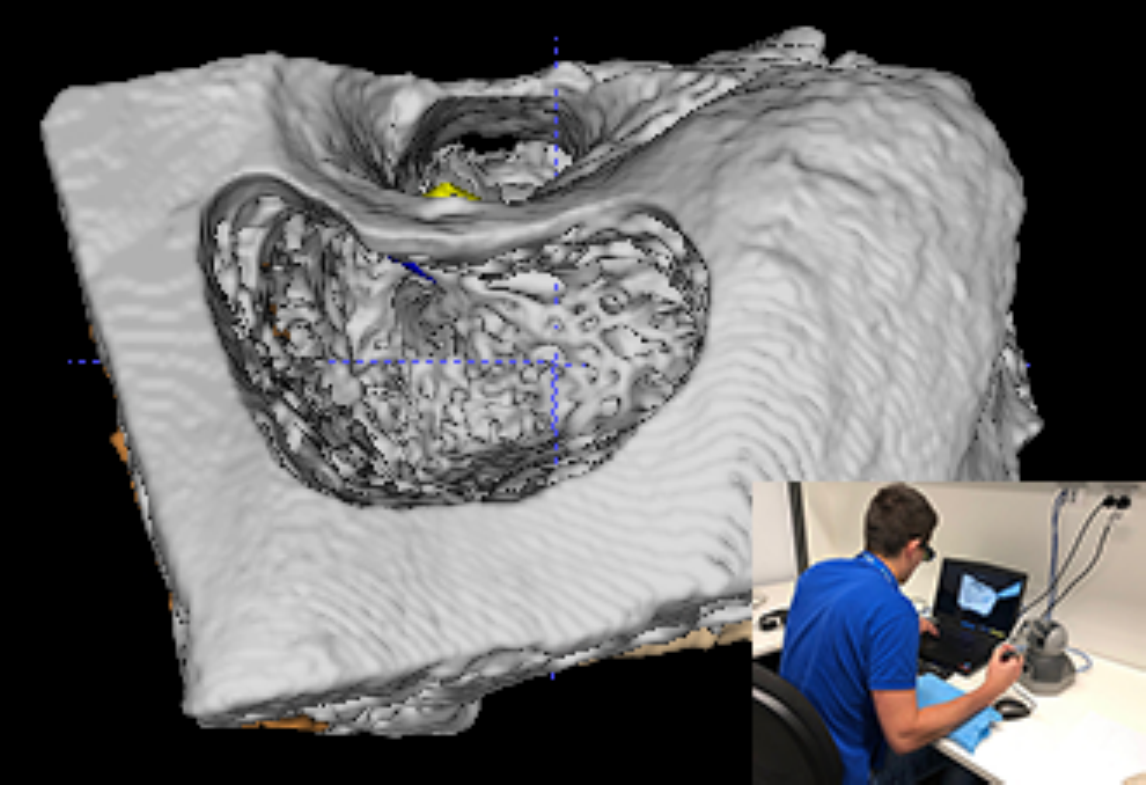}
  \caption{Processed temporal bone image after the cortical mastoidectomy (CM) surgery (bottom right is one surgeon performing a CM surgery in the VRTBS simulator).}
  \label{image:temporal}
\end{figure}

The Cortical Mastoidectomy (CM) dataset is collected with the help of expert ear surgeons performing this operation on the Virtual Reality Temporal Bone Surgery (VRTBS) simulator \cite{ma2017adversarial}. The VRTBS simulator was developed as a platform for temporal bone surgery training, including CM. Expert surgeons can record their surgeries in the simulator so that trainees can learn from them. Trainees can also practise performing surgeries repetitively in the simulator before they achieve expertise. Long time series of the surgical drilling bit are recorded as voxel positions in the 3D space. Fig. \ref{image:temporal} is an example of a temporal bone after the completion of a CM surgery on the simulator \cite{py06nimg}. The vacant region in the center is the drilled part of the mastoid from a temporal bone. Data preprocessing is implemented for better distance measurement because surgical time series collected in the simulator are usually constrained with noise. First, all consecutive and duplicated elements are deleted so that no repeated positions are recorded. Second, we also delete positions without the removal of the mastoid so that all changes of positions in time series are effective actions. In the end, remaining time series with varying lengths are saved for further distance measurement. \cite{tan2019time} proposed processing methods to deal with varying-length time series, such as the uniform scaling, the prefix and suffix padding, etc. We do not include them because (1) there is no noticeable improvement and (2) DTW and its derivatives are able to measure distances between time series with varying lengths.

Surgeons remove the mastoid part by part so that time series with large discontinuities are recorded. Surgeons have a variety of ways to remove mastoid air cells in different styles. Time series of the surgical drilling bit in every surgery are different from each other from the global point of view. However, some parts of them are analogous to each other from the local point of view because surgeons tend to remove parts of the mastoid in their own way when they perform CM surgeries. As the surgery goes on, there are more and more stochastic actions of removing the mastoid out of the temporal bone, which is unavoidable to be recorded as stochastic elements in time series. These stochastic elements can impede the discovery of surgeries from the same surgeon severely. In order to alleviate this problem, we truncate the first $1/5$ of all time series to build the CM dataset. 21 surgeries are collected from 7 surgeons on the same temporal bone, with each surgeon performing 3 surgeries in their unique style. Therefore, there are 21 time series pairs from the same surgeon and 189 pairs from different surgeons in total. Surgical time series from the same surgeon tend to share smaller distances between each other than those from different surgeons.

\begin{figure*}
\centering
  \includegraphics[width=\textwidth]{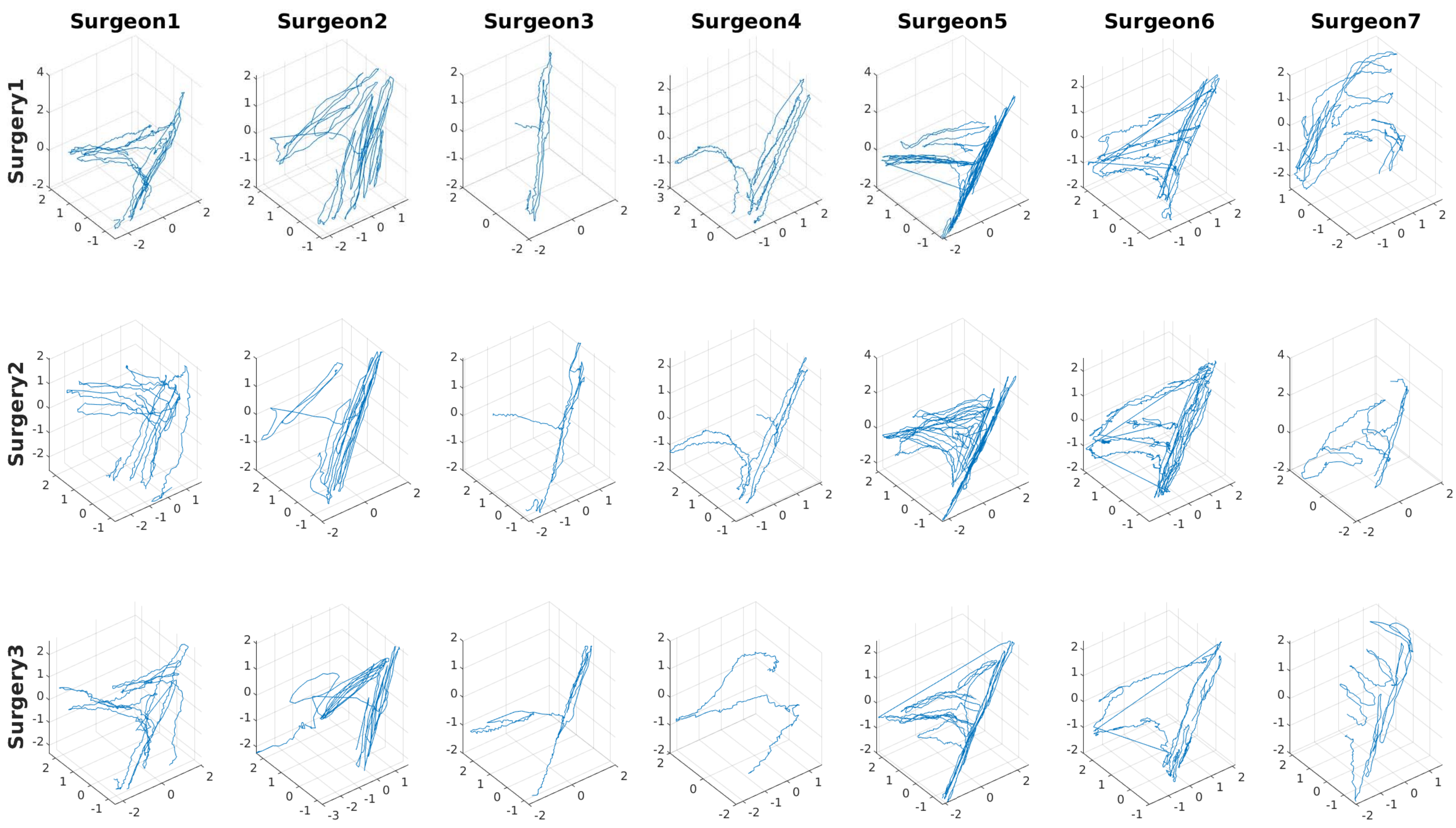}
  \caption{Cortical mastoidectomy surgery time series performed by seven surgeons with each performing three surgeries.}
  \label{image:surgeon}
\end{figure*}

To our knowledge, the CM dataset is a very challenging benchmark dataset. Surgeries from the same surgeon can still be different from each other from a global point of view (Fig. \ref{image:surgeon}). Our goal is not to propose a novel algorithm and identify every surgery from every surgeon without any mistake (we believe that no algorithms can successfully do it right now). We provide this benchmark dataset to help researchers explore the possibility of distance measurement for time series with large discontinuities. There are some significant characteristics of the CM dataset:

\begin{itemize}
\item All surgical time series are collected from expert surgeons in the VRTBS simulator. It is different from those time series collected from sensors or other professional tools because there are not only noise but also unavoidable stochastic actions in them, which makes it more challenging to measure distances between these time series pairs.
\item Apart from the noise and unavoidable stochastic actions in the CM dataset, there are large discontinuities in every surgical time series. The detection of consecutive elements with large discontinuities is crucial for measuring distances between them.
\item All surgical time series collected in the CM dataset are 3D time series. while most open datasets are composed of 1D (e.g., stock price, ECD) or 2D (e.g., GPS trajectories) time series.
\end{itemize}

\subsection{Open datasets}

There are not a lot of open datasets where time series own large discontinuities. This should be one reason why few research addressed the issue of measuring distances between them before. However, it is still important to discuss about this issue as they are common in some scenarios, such as surgical procedures and human activity. We will release the CM dataset as a supplement to existing datasets. It will be beneficial to researchers who want to do further research on measuring distances between time series with large discontinuities.

In order to validate the segmented pairwise distance (SPD) algorithm, we should find some other open datasets and modify them if necessary. After thorough (may not be complete) searching throughout the UCR Archive and other online data resources, two datasets are used for our experiments after limited modification.

\subsubsection{Activity Recognition dataset}
The first dataset is the Activity Recognition (AG) dataset \cite{casale2012personalization}. AG is collected from wearable accelerometers mounted on chests of 15 participants performing 7 activities, such as standing, walking, going updown stairs, etc. It provides challenges for identification and authentication of people using motion patterns.

Although there are 7 different activities in the recorded time series of every participant, there is still no large discontinuities in time series. In order to build time series with large discontinuities as the benchmark dataset, we concatenate different activities of the same participant to build new time series for everyone. The activity standing and walking are selected because their patterns are quite different from each other in fluctuation amplitude and frequency. For every participant, two new time series are built as $(S_1, W_1, S_2, W_2)$ and $(W_3, S_3, W_4, S_4)$, where $S$ stands for standing and $W$ stands for walking. They are randomly selected from subsequences of time series representing the same participant with the same length (500 elements in each subsequence and 2000 in total). We partition 15 participants into 5 sub-datasets so that there are 3 participants in each experiment in AG (from $AR_1$ to $AR_5$ in Table \ref{table:experiment}), with each participant having two newly-built time series.

\subsubsection{Indoor User Movement dataset}
The second dataset is the Indoor User Movement (IUM) dataset \cite{bacciu2014experimental}. IUM contains patterns of user movements in real-world office environments from time series generated by a Wireless Sensor Network (WSN), comprising 5 sensors: 4 anchors deployed in the environment and 1 mote worn by the user. Target data in IUM consists in a class label indicating whether the user's time series will lead to a room change or not. In particular, the target class 1 is associated to the location changing movements (156 sequences), while the target class -1 is associated to the location preserving movements (158 sequences).

For the same reason mentioned above, we need to build new time series to evaluate the performance of SPD. Every movement is recorded by 4 anchors in IUM. We can concatenate them together and add constant values in the second and fourth subsequences to create large discontinuities. We select 26 time series from IUM non-repetitively every time, with 13 from class 1 and the other 13 from class -1. As a result, the IUM dataset is partitioned into 12 sub-datasets (from $IUM_1$ to $IUM_{12}$ in Table \ref{table:experiment}).

\section{Experimental Results and Analysis}
In this section, we will compare and analyze the performance of SPD-embedded algorithms with their corresponding distance-based ones on open datasets and the cortical mastoidectomy (CM) dataset collected by expert surgeons.

\subsection{Silhouette index}
To evaluate the performance of SPD-embedded algorithms along with corresponding distance-based ones, we should test if they measure distances between time series reasonably. The internal cluster criteria are built to evaluate the performance of clustering algorithms, which compare average within-cluster distances and between-cluster distances obtained by different algorithms \cite{jeong2011weighted}. There are a group of validity indices to interpret and validate distance measurement of clustering algorithms, including Silhouette index, Dunn index, Davies–Bouldin index, etc. In this paper, the Silhouette index (SI) is used as an evaluation metric for our task \cite{rousseeuw1987silhouettes}. Equations \eqref{equ:s1}-\eqref{equ:s5} define SI, where $t$ are time series in the dataset $\textbf{D}_t$. We can calculate $SI(t_i)$ for every time series $t_i$ and the overall $SI(\textbf{D}_t)$ can be an estimation of the performance of distance-based algorithms on the dataset. $a(t_i)$ is the average pairwise distance between $t_i$ and any other time series in the same cluster while $b(t_i)$ is the average pairwise distance between $t_i$ and any time series in the neighbouring cluster. SI are in range of $[-1, 1]$ and high SI means appropriate distance measurement for clustering the dataset. 

\begin{equation}
a(t_i)=\frac{1}{|C_{t_i}|-1}\sum_{j \in C_{t_i}, i \neq j}D(t_i,t_j)
  \label{equ:s1}
\end{equation}
\begin{equation}
b(t_i)=\min_{k \neq i}\frac{1}{|C_{t_k}|}\sum_{t_j \in C_{t_k}}D(t_i,t_j)
  \label{equ:8s2}
\end{equation}
\begin{equation}
SI(t_i)=\frac{b(t_i)-a(t_i)}{\max (a(t_i),b(t_i))}, \; if \; |C_{t_i}|>1
  \label{equ:s3}
\end{equation}
\begin{equation}
SI(t_i)=0, \; if \; |C_{t_i}|=1
  \label{equ:s4}
\end{equation}
\begin{equation}
SI(\textbf{D}_t)=\frac{1}{|\textbf{D}_t|}\sum_{t_i \in \textbf{D}_t}SI(t_i)
  \label{equ:s5}
\end{equation}

\subsection{Experimental algorithms}
\begin{table*}
\caption{Experimental results on all datasets measured by SI (SI in range of $[-1, 1]$; CM: cortical mastoidectomy dataset; AG: activity recognition dataset; IUM: indoor user movement dataset).}
\label{table:experiment}
\begin{center}
  \begin{tabular}{c|cc|cc|cc|cc|cc}
     \toprule
     & DTW & \textbf{SDTW} & CIDTW & \textbf{SCIDTW} & DDTW & \textbf{SDDTW} & $\textrm{WDTW}_{0.01}$ & $\textbf{SWDTW}_{0.01}$ & $\textrm{WDDTW}_{0.01}$ & $\textbf{SWDDTW}_{0.01}$ \\
    \midrule
    CM & $0.054$ & $0.139$ & $0.156$ & $\textbf{\emph{0.226}}$ & $-0.234$ & $-0.044$ & $-0.627$ & $0.009$ & $-0.673$ & $-0.094$ \\
    \midrule
    $\textrm{AR}_1$ & $0.226$ & $0.260$ & $0.245$ & $\textbf{\emph{0.309}}$ & $0.220$ & $0.222$ & $-0.198$ & $0.223$ & $-0.171$ & $0.038$ \\
    $\textrm{AR}_2$ & $0.264$ & $0.406$ & $0.355$ & $\textbf{\emph{0.415}}$ & $0.186$ & $0.182$ & $-0.216$ & $0.138$ & $-0.208$ & $0.056$ \\
    $\textrm{AR}_3$ & $0.291$ & $0.488$ & $0.376$ & $\textbf{\emph{0.504}}$ & $0.248$ & $0.277$ & $-0.199$ & $0.315$ & $-0.174$ & $0.154$ \\
    $\textrm{AR}_4$ & $0.280$ & $0.449$ & $0.351$ & $\textbf{\emph{0.539}}$ & $0.161$ & $0.117$ & $-0.193$ & $0.203$ & $-0.147$ & $-0.063$ \\
    $\textrm{AR}_5$ & $0.091$ & $0.459$ & $0.180$ & $\textbf{\emph{0.511}}$ & $0.060$ & $0.144$ & $-0.168$ & $0.196$ & $-0.144$ & $-0.064$ \\
    Overall & $0.230$ & $0.412$ & $0.301$ & $\textbf{\emph{0.456}}$ & $0.175$ & $0.188$ & $-0.195$ & $0.215$ & $-0.169$ & $0.024$ \\
    \midrule
    $\textrm{IUM}_1$ & $0.020$ & $0.140$ & $0.013$ & $\textbf{\emph{0.169}}$ & $0.023$ & $0.026$ & $0.023$ & $0.137$ & $0.028$ & $0.059$ \\
    $\textrm{IUM}_2$ & $0.031$ & $0.155$ & $0.070$ & $\textbf{\emph{0.163}}$ & $0.022$ & $0.071$ & $0.040$ & $0.145$ & $0.031$ & $0.069$ \\
    $\textrm{IUM}_3$ & $0.013$ & $0.232$ & $0.017$ & $0.143$ & $-0.017$ & $0.032$ & $0.031$ & $\textbf{\emph{0.264}}$ & $0.006$ & $0.110$ \\
    $\textrm{IUM}_4$ & $0.076$ & $0.295$ & $0.074$ & $0.318$ & $-0.031$ & $0.043$ & $0.112$ & $\textbf{\emph{0.325}}$ & $0.035$ & $0.095$ \\
    $\textrm{IUM}_5$ & $0.059$ & $0.385$ & $0.069$ & $0.348$ & $0.007$ & $0.051$ & $0.129$ & $\textbf{\emph{0.424}}$ & $0.071$ & $0.201$ \\
    $\textrm{IUM}_6$ & $0.093$ & $0.123$ & $0.114$ & $\textbf{\emph{0.166}}$ & $-0.010$ & $0.037$ & $0.120$ & $0.109$ & $0.051$ & $0.066$ \\
    $\textrm{IUM}_7$ & $0.136$ & $\textbf{\emph{0.478}}$ & $0.127$ & $0.472$ & $-0.061$ & $0.087$ & $0.199$ & $0.451$ & $0.069$ & $0.113$ \\
    $\textrm{IUM}_8$ & $-0.043$ & $0.112$ & $-0.041$ & $\textbf{\emph{0.180}}$ & $-0.084$ & $0.043$ & $0.116$ & $0.135$ & $0.078$ & $0.128$ \\
    $\textrm{IUM}_9$ & $-0.019$ & $0.366$ & $-0.012$ & $\textbf{\emph{0.403}}$ & $-0.082$ & $0.093$ & $0.111$ & $0.338$ & $0.076$ & $0.150$ \\
    $\textrm{IUM}_{10}$ & $0.024$ & $0.212$ & $0.028$ & $\textbf{\emph{0.246}}$ & $-0.058$ & $0.052$ & $0.145$ & $0.239$ & $0.080$ & $0.104$ \\
    $\textrm{IUM}_{11}$ & $0.001$ & $0.082$ & $0.004$ & $0.086$ & $-0.045$ & $0.013$ & $0.107$ & $\textbf{\emph{0.127}}$ & $0.066$ & $0.074$ \\
    $\textrm{IUM}_{12}$ & $0.083$ & $0.233$ & $0.081$ & $0.161$ & $-0.056$ & $0.130$ & $0.163$ & $\textbf{\emph{0.241}}$ & $0.078$ & $0.185$ \\
    Overall & $0.040$ & $0.234$ & $0.045$ & $0.238$ & $-0.033$ & $0.057$ & $0.108$ & $\textbf{\emph{0.245}}$ & $0.056$ & $0.113$ \\
    \bottomrule
  \end{tabular}
\end{center}
\end{table*}

We select 5 distance-based algorithms, namely DTW, CIDTW, DDTW, WDTW and WDDTW mentioned in Section \ref{section2}. We chose DTW as an example in Section \ref{section3} to exhibit how SPD can be embedded in distance-based algorithms. Here are other four algorithms which are to be compared in following experiments.

\subsubsection{CIDTW}
Complexity invariance was proposed as a supplement to the invariance family including amplitude invariance, local scaling invariance, uniform scaling invariance, phase invariance, occlusion invariance, and their combinations \cite{batista2011complexity}. The complexity invariance was achieved by the introduction of a correction factor $CF$ for existing distance measures, obtained by the complexity estimate $CE$. It can be embedded in DTW as the complexity-invariant DTW (CIDTW) algorithm defined by \eqref{equ:ci1}-\eqref{equ:ci3}.

\begin{equation}
  CIDTW(\bm{A}, \bm{B}) = DTW(\bm{A}, \bm{B}) \times CF(\bm{A}, \bm{B})
  \label{equ:ci1}
\end{equation}
\begin{equation}
  CF(\bm{A}, \bm{B}) = \frac{max(CE(\bm{A}), CE(\bm{B}))}{min(CE(\bm{A}), CE(\bm{B}))}
  \label{equ:ci2}
\end{equation}
\begin{equation}
 CE(\bm{A}) = \sqrt{\sum_{i=1}^{n-1}(a_i-a_{i+1})^2}
  \label{equ:ci3}
\end{equation}

\subsubsection{DDTW} 
The derivative DTW (DDTW) algorithm aligns time series considering high level features of shape. It obtains information about shapes by considering the first derivative of time series \cite{keogh2001derivative}. DDTW preprocesses time series using \eqref{equ:ddtw}, where the undefined elements $a_1^{D}$ and $a_n^{D}$ are obtained by $a_1^{D}=a_2^{D}$ and $a_n^{D}=a_{n-1}^{D}$. Distances between preprocessed time series are then calculated by DTW.

\begin{equation}
  a_i^{D} = \frac{(a_i-a_{i-1})+(a_{i+1}-a_{i-1})/2}{2}, \; 1<i<n
  \label{equ:ddtw}
\end{equation}

\subsubsection{WDTW}
The phase difference is also a common problem because DTW provides non-linear alignments, which is regarded as the phase invariance problem in the invariance family. \cite{jeong2011weighted} proposed the weighted DTW (WDTW) algorithm to penalize elements with larger phase difference using \eqref{equ:wdtw1}-\eqref{equ:wdtw2}, in order to achieve minimum distance distortion caused by outliers. Equation \eqref{equ:wdtw3} is the modified logistic weight function (MLWF) proposed to systematically assign weights as a function of phase difference. $g$ is the penalty coefficient for phase difference. There is no guarantee that SPD can improve the performance of WDTW for any $g$ on any dataset, but we found positive results in most experiments. $g=0.01$ is selected as the penalty for WDTW shown in Table \ref{table:experiment}.

\begin{equation}
\begin{split}
  D(1{:}n_1,1{:}n_2) = w_{n_1-n_2}d(a_{n_1},b_{n_2}) + min[D(1{:}(n_1-1),\\
  1{:}(n_2-1)), D(1{:}(n_1-1),1{:}n_2),D(1{:}n_1,1{:}(n_2-1))]
\end{split}
\label{equ:wdtw1}
\end{equation}
\begin{equation}
  D(1{:}1,1{:}1) = w_{1-1}d(a_1,b_1)
  \label{equ:wdtw2}
\end{equation}
\begin{equation}
  w_i = \frac{w_{max}}{1+\exp{(-g(i-n_c))}}
  \label{equ:wdtw3}
\end{equation}

\subsubsection{WDDTW}
The penalty for phase difference can be extended to variants of DTW, including DDTW. The weighted version of DDTW (WDDTW) was then proposed in \cite{jeong2011weighted}. We use the same $g$ for WDDTW in all experiments.

\subsection{Results and analysis}
We validate advantages of SPD by selecting 5 distance-based algorithms with their corresponding SPD-embedded versions to measure pairwise distances on the CM, AR and IUM datasets, respectively. All odd columns are results of existing distance-based algorithms and all even columns are those of corresponding SPD-embedded algorithms. The quantile of sorted distance distribution is insensitive when it is in range of $[0.9, 0.99]$ in most scenarios. We set the quantile to be $0.99$ in all experiments based on a priori knowledge of datasets.

On the CM dataset, all SPD-embedded algorithms perform better than corresponding distance-based ones, with overall improvement of $\textbf{\emph{0.312}}$ in average measured by SI. DDTW, WDTW and WDDTW all perform poorly on the CM dataset. Although their corresponding SPD-embedded algorithms improve much based on their poor performance, they are still worse than DTW. The poor performance of DDTW implies that high level features of shape extracted by the first derivative of time series should impede the measurement of their pairwise distances in the CM dataset. The poor performance of WDTW and WDDTW implies that we should tolerate the phase difference between time series when measuring their pairwise distances in the CM dataset. SCIDTW performs the best on the CM dataset, with CIDTW the second and SDTW the third. SDTW does not defeat CIDTW but CIDTW can be additionally improved by our proposed SPD as the champion SCIDTW on the CM dataset. SDTW improves the performance of DTW by $157 \%$ and SCIDTW improves the performance of CIDTW by $45 \%$, respectively.

On the AR dataset, most SPD-embedded algorithms perform better than corresponding distance-based ones on sub-datasets, with overall improvement of $\textbf{\emph{0.19}}$ in average measured by SI. DDTW performs slightly worse than DTW, while WDTW and WDDTW perform poorly on the AR dataset. Although their corresponding SPD-embedded algorithms improve much based on their poor performance, they are still worse than DTW. The poor performance of DDTW implies that it is not necessary to extract high level features of shape by the first derivative of time series when measuring their pairwise distances in the AR dataset. The poor performance of WDTW and WDDTW implies that we should also tolerate the phase difference between time series in the AR dataset. SCIDTW performs the best on the AR dataset, with SDTW the second. SDTW defeats CIDTW on the AR dataset. Both SCIDTW and SDTW perform much better than all other algorithms. SDTW improves the performance of DTW by $79 \%$ and SCIDTW improves the performance of CIDTW by $51 \%$, respectively.

On the IUM dataset, most SPD-embedded algorithms perform better than corresponding distance-based ones on sub-datasets as well, with overall improvement of $\textbf{\emph{0.134}}$ in average measured by SI. However, almost all distance-based algorithms perform badly on the IUM dataset. It is quite necessary to use SPD-embedded algorithms in order to improve the performance of corresponding distance-based ones in measuring pairwise distances of time series in the IUM dataset. SWDTW performs the best on the IUM dataset, with SCIDTW the second and SDTW the third. SDTW, SCIDTW, and SWDTW improves the performance of DTW, CIDTW, and WDTW by $485 \%$, $429 \%$, and $127 \%$, respectively. The top 3 algorithms share very close performance to each other, which is one main reason why we do experiments on randomly-selected sub-datasets. We can clearly see the distribution of best performance on these sub-datasets when overall results are close to each other.

In conclusion, all algorithms perform quite differently from each other on different datasets. SPD-embedded algorithms can help improve the performance of corresponding distance-based algorithms dominantly on every dataset, even when distance-based ones perform very badly. DTW is a widely used algorithm, which is hard to beat by its derivative algorithms (CIDTW, DDTW, WDTW and WDDTW). CIDTW performs the best among 5 distance-based algorithms, which shows the importance of achieving complexity invariance when measuring distances between time series with large discontinuities. Moreover, none of these distance-based algorithms are ubiquitously well-performing ones and they can only perform well in their specific scenarios. The poor performance of SWDDTW on all datasets implies that complicated algorithms can not make sure of good performance. It may not be necessary to learn shape features of time series or consider the phase invariance when measuring their pairwise distances all the time. It is always necessary to obtain a priori knowledge of scenarios in order to select suitable algorithms. In this scenario where all time series own large discontinuities, SCIDTW performs the best, followed by SDTW.

\section{Conclusions and Future Work}
This paper proposes a new algorithm, the segmented pairwise distance (SPD), to measure distances between time series with large discontinuities, which are common in many scenarios. SPD is orthogonal to distance-based algorithms and can be embedded in them. We validate advantages of SPD-embedded algorithms over corresponding distance-based ones on both open datasets and our collected cortical mastoidectomy (CM) dataset. We provide the potential of distance measurement with SPD-embedded algorithms in more challenging scenarios. In the near future, we plan to (1) find an intelligent method to decide the segmentation threshold for SPD on different datasets and (2) consider the extension of SPD to surgical time series identification, human activity recognition and other challenging tasks.

\section*{Acknowledgment}
We truly appreciated great cooperation with $7$ anonymous surgeons at the Royal Victorian Eye and Ear Hospital, who helped us perform and validate CM surgeries in the VRTBS simulator. This research was supported by the Melbourne Research Scholarship.

\bibliographystyle{IEEEtran}
\bibliography{IEEEabrv,IJCNN20}

\end{document}


\begin{figure}
\centering
    \begin{subfigure}{0.25\textwidth}
      \includegraphics[width=\textwidth]{sup/example11.pdf}
      \caption{DTW=22}
    \end{subfigure}%
    \begin{subfigure}{0.25\textwidth}
      \includegraphics[width=\textwidth]{sup/example12.pdf}
      \caption{SDTW=2}
    \end{subfigure}%
    \begin{subfigure}{0.25\textwidth}
      \includegraphics[width=\textwidth]{sup/example13.pdf}
      \caption{MHD=0}
    \end{subfigure}%
\caption{Example of Similarity Evaluation for Time Series with Large Discontinuity using DTW, SDTW, and MHD}
\Description{The first sequence is (4, 5, 6, 1, 2, 3, 7, 8, 9). The second sequence is (1, 2, 3, 7, 8, 9, 4, 6, 5). From the local point of view, two sequences are very similar to each other but not exactly the same. SDTW outpforms DTW and MHD when analyzing local spatio-temporal similarity between the two sequences.}
\vspace{2cm}
\end{figure}

\begin{figure}
  \centering
    \begin{subfigure}{0.4\textwidth}
      \includegraphics[width=\textwidth]{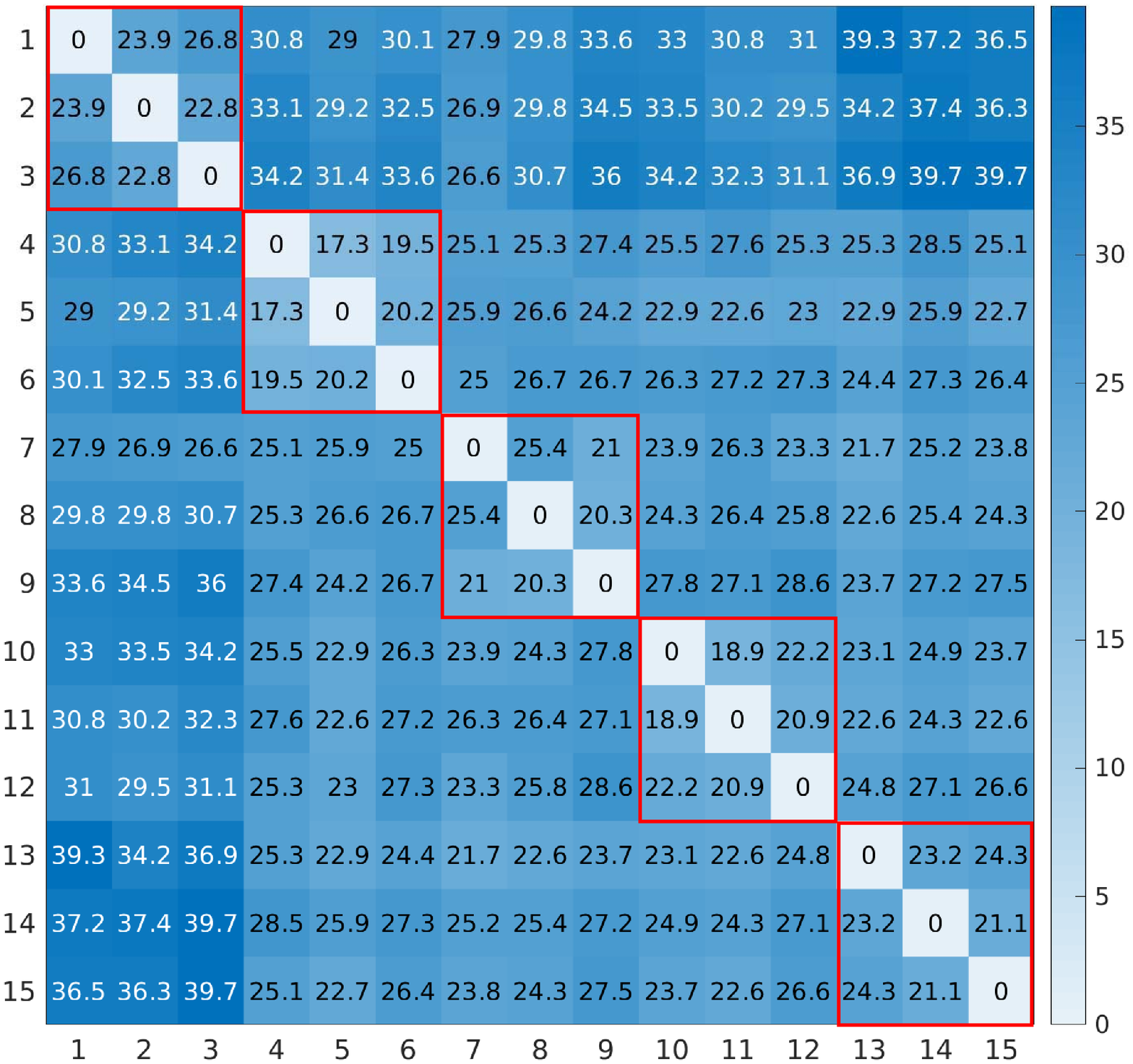}
      \caption{DTW}
    \end{subfigure}%
    \begin{subfigure}{0.4\textwidth}
      \includegraphics[width=\textwidth]{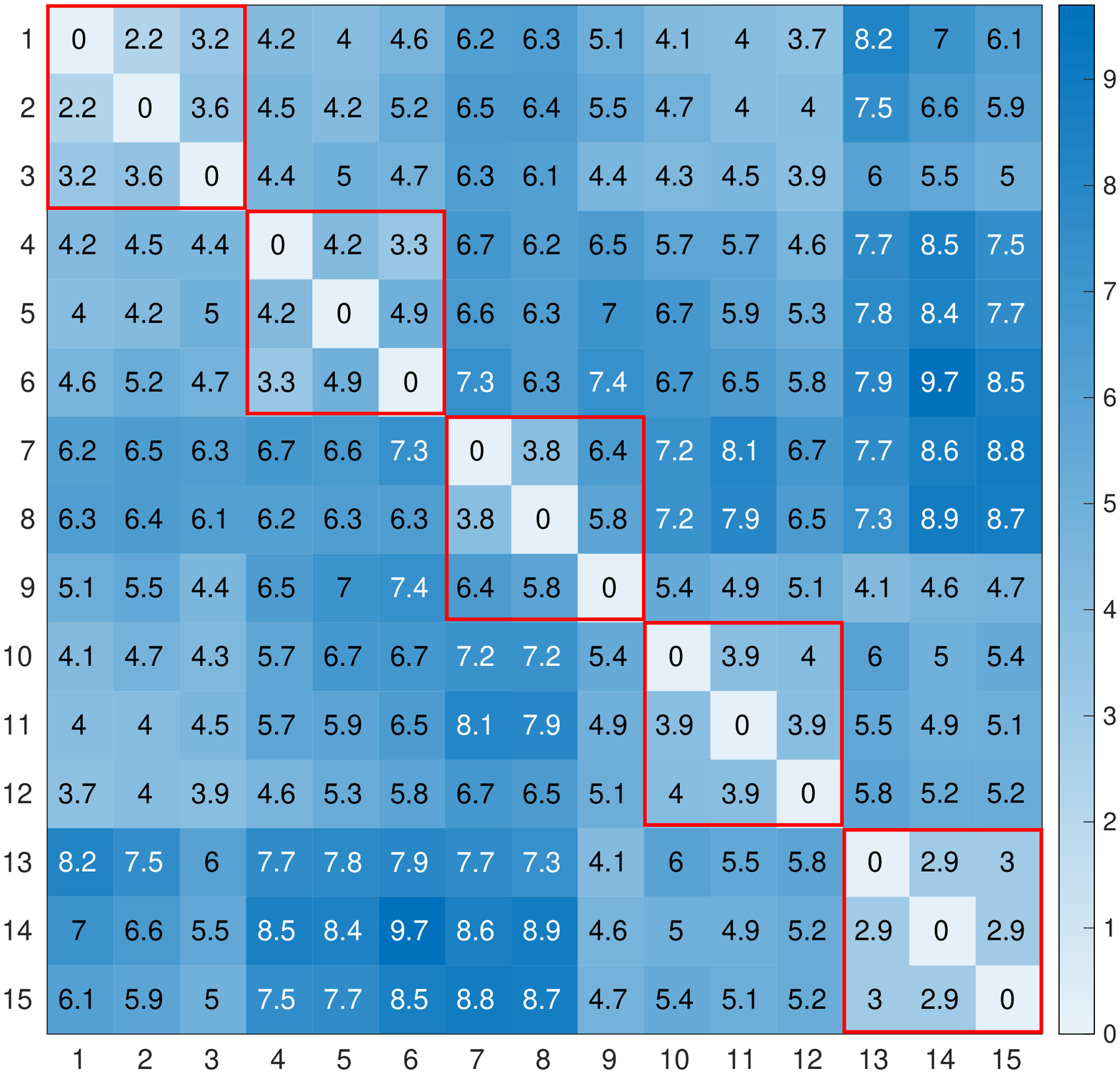}
      \caption{MHD}
    \end{subfigure}
  \caption{Comparison of Distance Matrices from Mastoidectomy Trajectories Calculated by DTW and MHD}
\end{figure}

\begin{figure}
  \centering
    \begin{subfigure}{0.4\textwidth}
      \includegraphics[width=\textwidth]{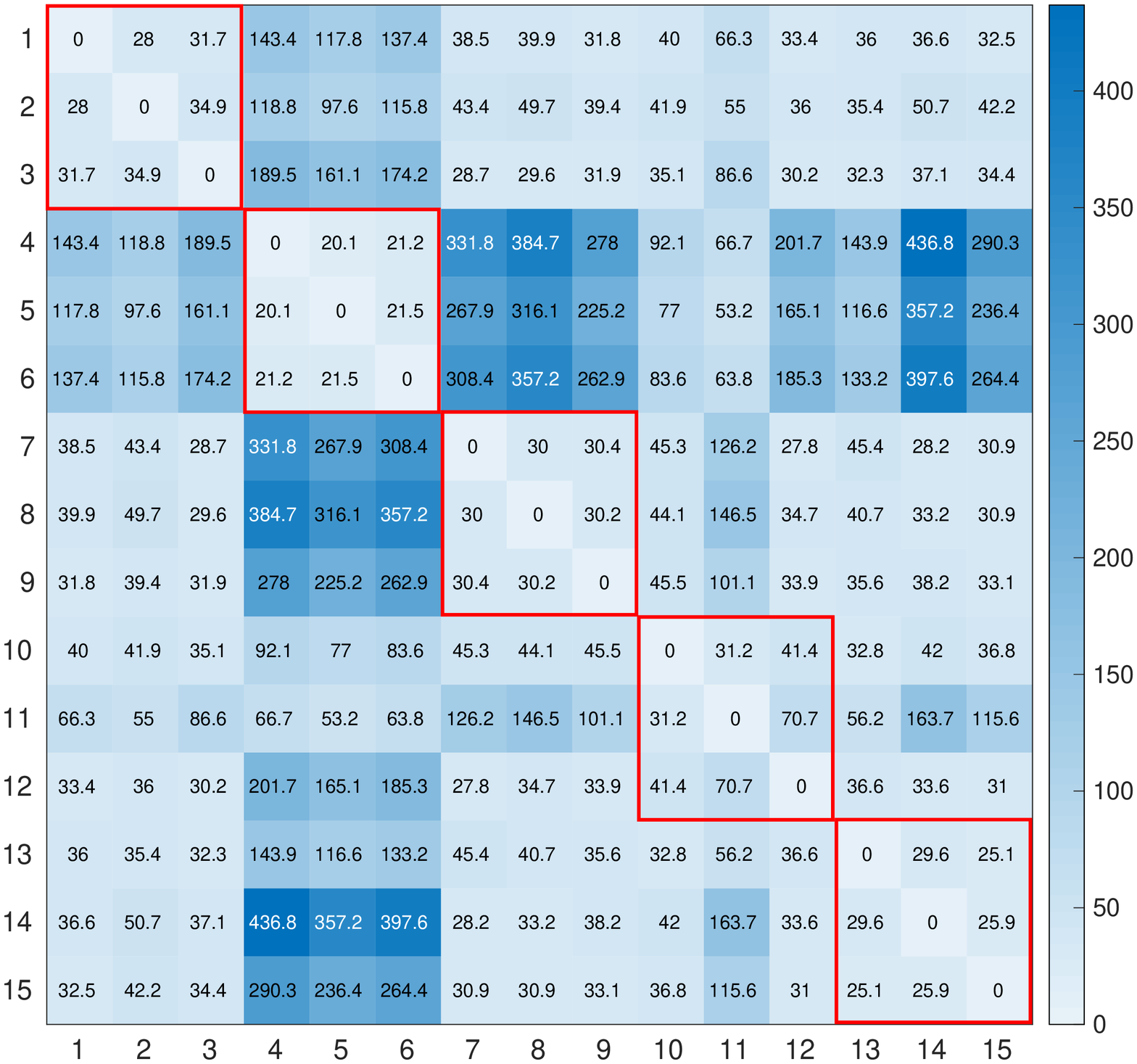}
      \caption{Max\_1}
    \end{subfigure}%
    \begin{subfigure}{0.4\textwidth}
      \includegraphics[width=\textwidth]{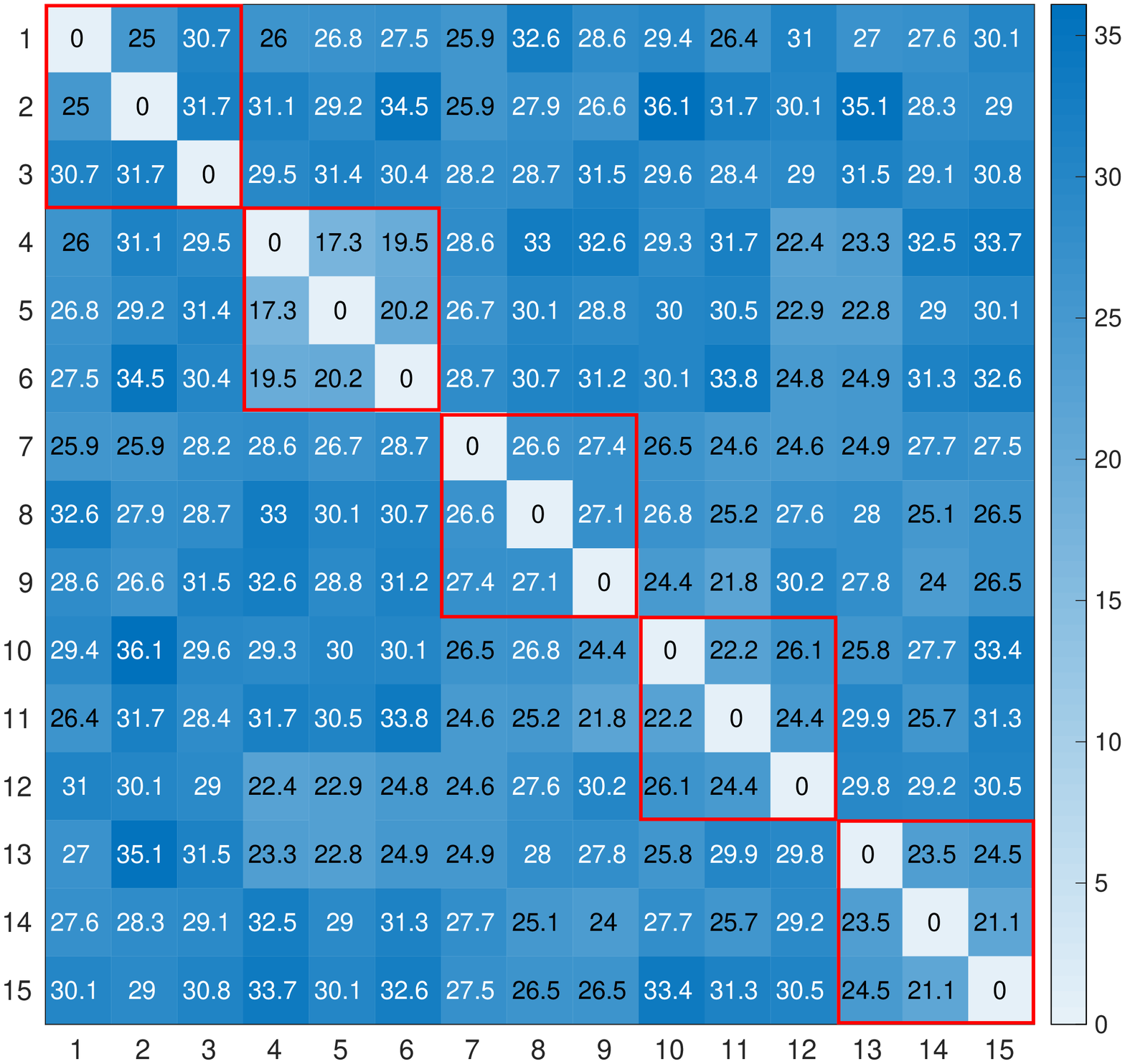}
      \caption{Min\_1}
    \end{subfigure}
    \hfill
    \begin{subfigure}{0.4\textwidth}
      \includegraphics[width=\textwidth]{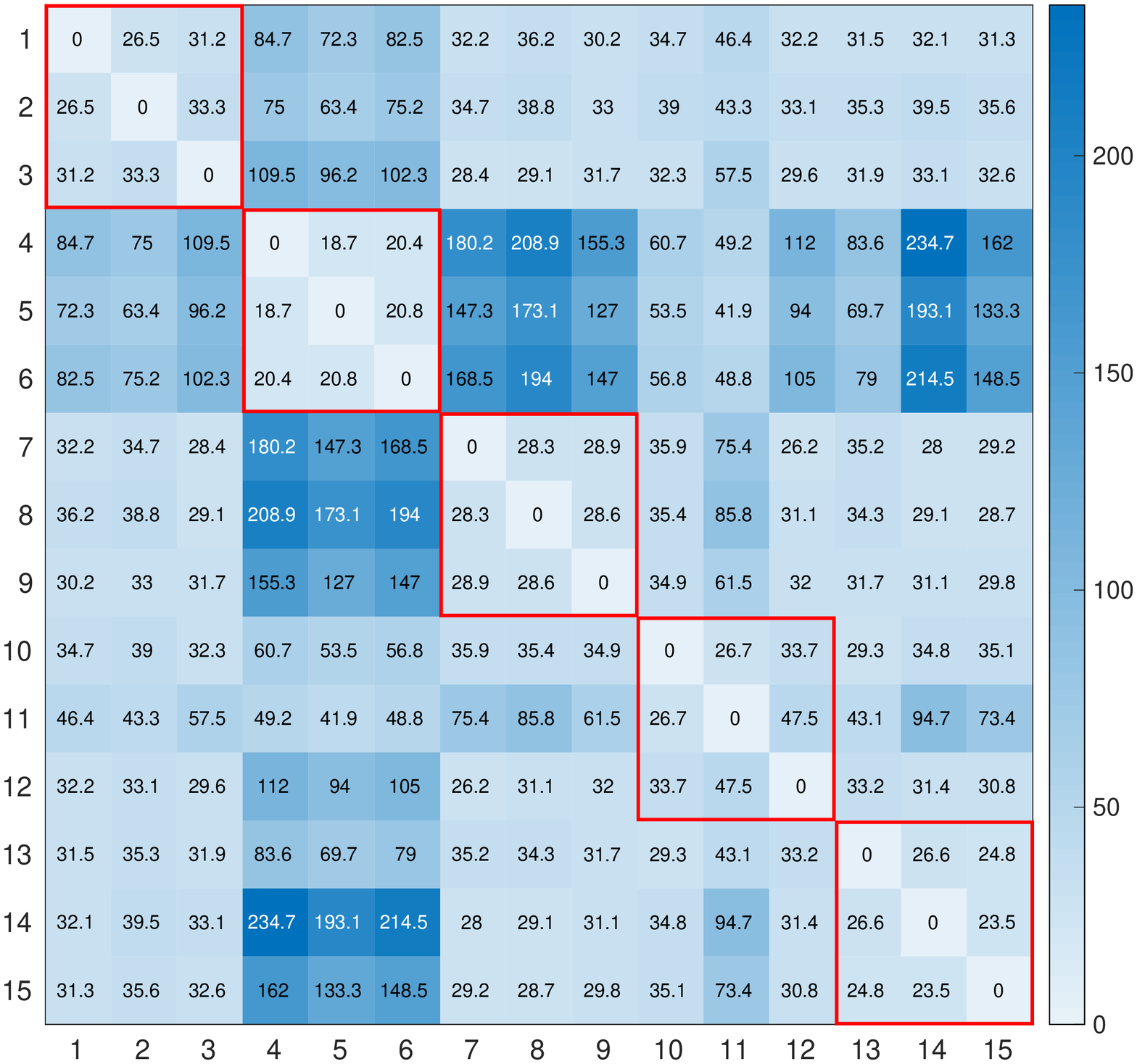}
      \caption{Mean\_1}
    \end{subfigure}%
    \begin{subfigure}{0.4\textwidth}
      \includegraphics[width=\textwidth]{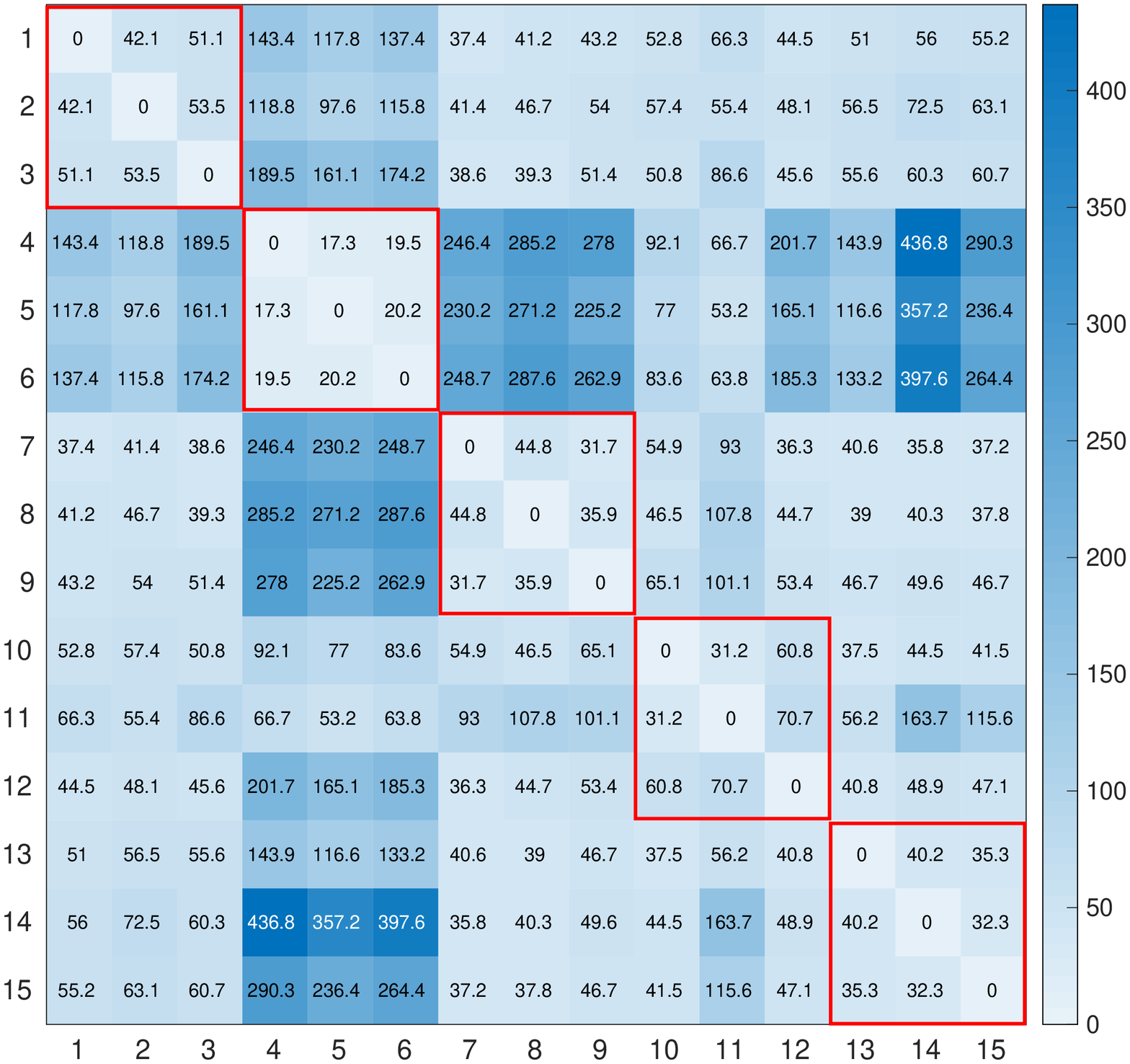}
      \caption{Max\_2}
    \end{subfigure}
    \hfill
    \begin{subfigure}{0.4\textwidth}
      \includegraphics[width=\textwidth]{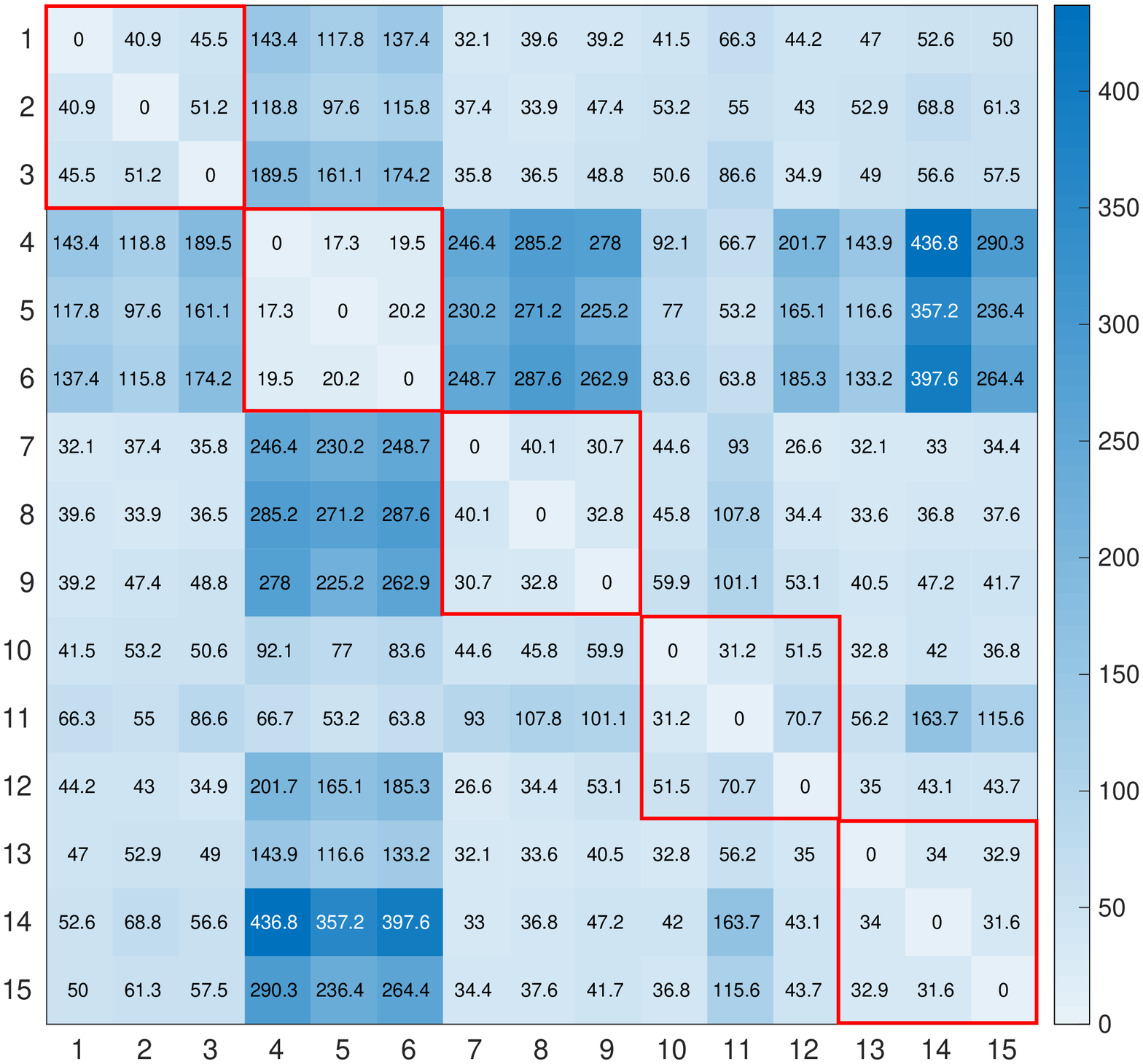}
      \caption{Min\_2}
    \end{subfigure}%
    \begin{subfigure}{0.4\textwidth}
      \includegraphics[width=\textwidth]{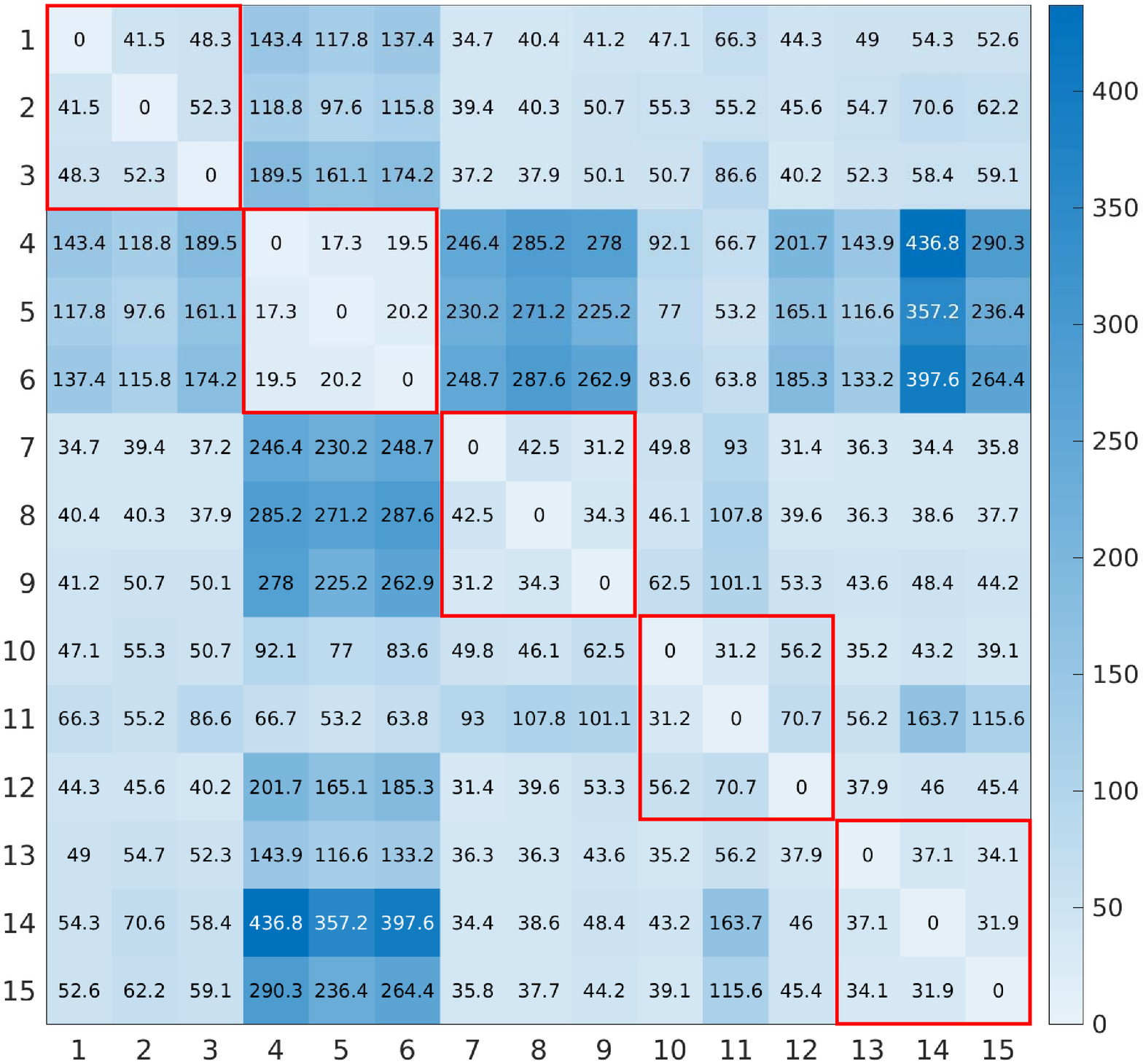}
      \caption{Mean\_2}
    \end{subfigure}
  \caption{Comparison of Distance Matrices from Mastoidectomy Trajectories Calculated by SDTW}
  \Description{(a), (b) and (c): distance matrices calculated by the first procedure; (d), (e) and (f): distance matrices calculated by the second procedure. The segmentation threshold is 48 voxels in (a), (b), (c), (d), (e) and (f).}
  \label{image:matrix}
\end{figure}